\documentclass[conference]{IEEEtran}

\usepackage[T1]{fontenc}
\usepackage{graphicx}
\usepackage{cite}
\usepackage{amsmath,amssymb,amsfonts}
\usepackage{textcomp}
\usepackage{xcolor}
\usepackage{soul}
\sethlcolor{lightgray}
\usepackage{paralist}
\usepackage{dirtytalk}
\usepackage{xurl}
\usepackage{booktabs}
\usepackage{hyperref}

\newcommand{\code}[1]{\textsf{\footnotesize#1}}
\newcommand{\opreturn}{\code{\textsc{op\!\_return}}}
\newcommand{\opcheckmultisig}{\code{\textsc{op\!\_checkmultisig}}}
\newcommand{\btc}{\textsc{btc}}

\IEEEoverridecommandlockouts 

\begin{document}

\title{No Country for Old Privacy: The Evolving Challenges of Anonymity in Bitcoin%
    ${}^*$
    \thanks{\hrule\vspace{2mm}${}^*$This is the author's version of a paper accepted to the 2nd IEEE International Conference on Distributed Ledger Technologies (ICDLT 2026).}
}

\author{
    \IEEEauthorblockN{Ben Hawkins}
    \IEEEauthorblockA{
    \textit{University of York}\\
    York, United Kingdom \\
    {\footnotesize\texttt{ben.hawkins.cs@gmail.com}}}
    \and
    \IEEEauthorblockN{Joshua Levett}
    \IEEEauthorblockA{
    \textit{University of York}\\
    York, United Kingdom \\
    {\footnotesize\texttt{joshua.levett@york.ac.uk}}}
    \and
    \IEEEauthorblockN{Siamak F.\ Shahandashti}
    \IEEEauthorblockA{
    \textit{University of York}\\
    York, United Kingdom \\
    {\footnotesize\texttt{siamak.shahandashti@york.ac.uk}}}
}

\maketitle

\begin{abstract}
  We present a longitudinal measurement study on the adoption of detectable, second-generation anonymisation protocols in the Bitcoin network, including CoinJoin, CoinSwap, CoinShuffle and Stealth Addresses. By implementing and refining a suite of heuristic filters, we identify over 5 million candidate CoinJoin and 23.3 million candidate CoinSwap transactions. Notably, the use of CoinShuffle was unexpectedly found to be closely aligned with the operation of coordinator-based CoinJoin services such as Wasabi and Whirlpool. Our analysis reveals consistently low adoption rates, with these protocols typically constituting less than 1\% of network transactions, and a sharp decline in detectable usage following key regulatory events. Furthermore, we find no evidence of standardised Stealth Address adoption, indicating a failure to converge on a common privacy standard. This study provides a comprehensive picture of a niche ecosystem whose on-chain visibility has been largely suppressed, strongly suggesting the migration of privacy-seeking users to less transparent and less detectable methods.
\end{abstract}

\begin{IEEEkeywords}
Bitcoin, Privacy, Anonymity, Privacy-Preserving Protocols,
CoinJoin, CoinSwap, CoinShuffle, Stealth Addresses
\end{IEEEkeywords}

\section{Introduction}
Bitcoin emerged in early 2009 after Nakamoto published a paper proposing a peer-to-peer electronic cash system built on a \say{blockchain}~\cite{nakamoto2008bitcoin}. This decentralised ledger records transactions securely, transparently, and tamper-resistantly without a central authority. Many early adopters were attracted by the prospect of owning and transferring digital currency with a high degree of anonymity.
In reality, the provided level of anonymity is quite limited. While users transact without revealing real identities, transactions involving the same addresses are trivially linkable. Furthermore, techniques such as transaction clustering, network analysis, and other heuristic methods~\cite{ReidH11,Ron2013-ez,Meiklejohn2013-lg,Harrigan2016-ak} can be used to link multiple addresses belonging to the same user with high confidence. 

In response to such de-anonymising techniques, a diverse range of privacy-preserving protocols began to appear. With a growing set of options available, privacy-conscious users were faced with the question of which protocol to adopt. Understanding what drives users' choices can inform the design of future protocols that not only better align with user preferences, but also encourage convergence on a smaller number of widely-used techniques, a necessity for building larger anonymity sets leading to stronger privacy.
Measuring how widely these protocols are actually adopted is therefore key to assessing the privacy users obtain in practice.

Such measurement has become more pressing as the Bitcoin ecosystem has evolved significantly: upgrades such as SegWit and Taproot have changed how transactions appear on-chain, user-friendly CoinJoin wallets have emerged, and regulatory pressure on privacy tools has intensified. How these changes have affected protocol adoption is not well understood.

\emph{Our Contributions.}
We study the adoption of second-generation privacy-preserving protocols in Bitcoin between 2016 and 2025 (blocks 430k--891k). Specifically:
\begin{itemize}
\item We measure the on-chain prevalence of CoinJoin, CoinSwap, CoinShuffle and Stealth Addresses using heuristic filters, extending Möser and Böhme's study~\cite{Moser2017-sb} by nine years, and release our implementation.
\item We find that adoption remained low, typically below 1\% of transactions, with CoinShuffle-style activity tracking the lifetime of coordinator-based CoinJoin services, and no evidence of standardised Stealth Address usage.
\item We relate adoption trends to major technical, market and regulatory events, observing a sharp decline in detectable activity around 2024, consistent with a migration to less detectable methods or reduced demand.
\end{itemize}

\section{Background}

A simple Bitcoin transaction specifies three elements: inputs locked by a previous \code{ScriptPubKey}, outputs containing a new \code{ScriptPubKey} defining how they can be spent, and the sender's unlocking data (\code{ScriptSig}). \code{ScriptSig} typically contains a digital signature and public key that prove ownership of the inputs and authorise their use, enabling the reallocation of coins.

Bitcoin scripts allow for more complex transactions, including hash/time-locked and multi-signature transactions. 
\emph{Hash-locked contracts} require the recipient to provide a secret (preimage of a hash) in order to spend the coins. The hash condition is written directly into the transaction’s locking script (\code{ScriptPubKey}). To spend the coins, the recipient must include the secret in their unlocking script (\code{ScriptSig}) and transactions are accepted by the network only if the hash matches. 
\emph{Time-locked contracts} require the passing of a certain amount of time after a transaction, before the coins can be spent. The lock condition is again specified in the output’s locking script, and nodes enforce it by rejecting any attempt to spend the coins too early. 
\emph{Hashed time-locked contracts} require a conjunctive or disjunctive combination of the two types of conditions above. 
\emph{Multi-signature (MultiSig) transactions} need multiple signatures in order to be spent. A standard transaction needs only a single signature from the sender (corresponding to the private key of the relevant input address), to authorise spending. However, an $m$-of-$n$ MultiSig transaction designates $n$ public keys and requires signatures from any $m$ of them to collectively authorise the transaction. 

In the early 2010s, centralised Bitcoin mixers were the dominant privacy solution. These worked by pooling together users’ coins, and redistributing them in a shuffled manner to obscure transaction links. However, because these services required users to hand over control of their funds, they were inherently prone to theft.
Möser and Böhme call these \emph{first-generation} protocols~\cite{Moser2017-sb}. While these protocols provided a valuable layer of transaction unlinkability, their centralised nature both contradicted the decentralised and trustless ethos underpinning Bitcoin, and also made them susceptible to surveillance, infiltration, and shutdown by law enforcement agencies. A second generation of protocols began to emerge, designed to preserve the privacy goals established by the first generation while avoiding their downsides. 

\subsection{CoinJoin and Derivatives}
Bitcoin requires each input unspent transaction output (UTXO) in a transaction to be signed separately, allowing multiple users to jointly create one transaction. This undermines the \say{common input ownership} heuristic, the assumption that all inputs in a transaction belong to one entity, used by de-anonymisation techniques such as clustering. With this method, users can obscure input-output address links if they all use similar quantities as outputs. 

\emph{CoinJoin.} 
Proposed by Maxwell~\cite{Maxwell2013coinjoin}, in CoinJoin, users first provide their input, output and change addresses, and individually sign the transactions. One user then combines the signatures and broadcasts the transaction to the network.

\emph{CoinShuffle.} 
Designed as an improvement to CoinJoin~\cite{Ruffing2014-em}, CoinShuffle provides measures against internal traceability (previously, any participant in the transaction would be able to trace the inputs and outputs), and also allows participants to verify the execution of the protocol, and blame misbehaving users if it is unsuccessful. CoinShuffle achieves this by using the Dissent protocol~\cite{Corrigan2010-ed}. After announcing the protocol, each participant generates a fresh ephemeral key pair, and broadcasts their public keys. After agreeing on a participant order, each participant $i$ receives $i-1$ ciphertexts containing layered encryptions of previous participants' output addresses in a randomised order, decrypts the outermost layer for all ciphertexts, encrypts their own output address under the remaining $n-i$ public keys, adds this layered ciphertext to the ciphertext set, shuffles the set, and sends the resulting $i$ ciphertexts to the next participant. Eventually, the final participant will have a shuffled list of $n$ participants' output addresses.
Figure~\ref{fig:coinshuffle-fairexchange}~(left) shows an example. 
Each participant can now verify that their address is in the final list. If they all are, the transaction is published, if not, the bad actor can be found if all participants publish their ephemeral private keys, allowing them to trace through the protocol and locate who deviated from the protocol. 

\subsection{Fair Exchange and CoinSwap}
\emph{Fair Exchange.} 
This protocol guarantees \emph{fairness} for two parties exchanging bitcoins. 
If either party aborts, both are fully refunded. Barber et al.\ gave the first Bitcoin-specific fair exchange protocol~\cite{Barber2012-yx}, 
which itself provides no privacy, but is widely used to build privacy-preserving protocols.

\begin{figure}[t]
    \centering
    \begin{minipage}{0.4\linewidth}
        \includegraphics[width=1\linewidth]{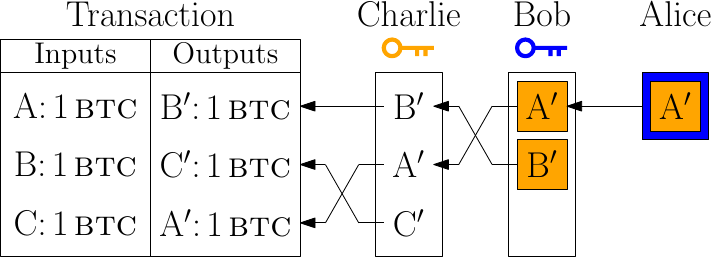}
    \end{minipage}%
    \quad\quad 
    \begin{minipage}{0.52\linewidth}
    \includegraphics[width=1\linewidth]{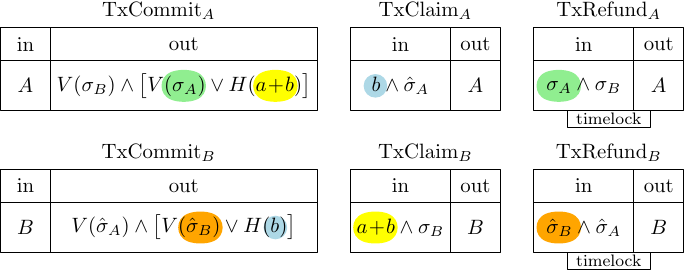} 
    \end{minipage}
    \caption{
        Left: A CoinShuffle protocol, where Alice has addresses $\mathrm{A,A'}$, Bob $\mathrm{B,B'}$, and Charlie $\mathrm{C,C'}$ (adapted from~\cite{Ruffing2014-em});
        Right: Fair Exchange transactions (adapted from~\cite{Barber2012-yx}).}
    \label{fig:coinshuffle-fairexchange}
\end{figure}

Suppose Alice and Bob want to swap coins without having to trust each other. 
First, they exchange signatures and establish refund transactions $\mathrm{TxRefund}_A$ and $\mathrm{TxRefund}_B$, both only claimable after time $T$. Next, Alice picks a random secret $a$ and sends it to Bob. Bob generates secret $b$ and sends $H(a+b)$ and $H(b)$ to Alice. A standard cut-and-choose protocol is used to ensure the hashes are in correct form with high probability. Bob then creates a transaction $\mathrm{TxCommit}_B$ redeemable by either Alice's and Bob's signatures, or Alice's signature and the preimage of $H(b)$. Alice then publishes $\mathrm{TxCommit}_A$ redeemable by either both Bob's and Alice's signatures, or Bob's signature and the preimage of $H(a+b)$. Now, for Bob to claim his coins from Alice, he must reveal $a+b$ in a transaction $\mathrm{TxClaim}_B$, in turn, allowing Alice to calculate $b$ and claim her coins from Bob through $\mathrm{TxClaim}_A$. If either party fails to complete the protocol, the refund transactions $\mathrm{TxRefund}_A$ and $\mathrm{TxRefund}_B$ ensure that both can reclaim their original coins after time $T$. 
Figure~\ref{fig:coinshuffle-fairexchange}~(right) shows the transactions involved.

Fair Exchange 
does not attempt to hide itself. It connects Alice and Bob with two transactions, making it trivially traceable, and hence undesirable.

\emph{CoinSwap.} This protocol adapts Fair Exchange to enable Alice to pay Bob without creating a visible link between their addresses on the blockchain~\cite{maxwell-coinswap}. This is achieved by introducing an intermediary, Carol. First, Carol and Bob create refund transactions $T_{0R}$ and $T_{1R}$ to ensure that if the protocol is aborted, Carol and Alice can recover their coins; $T_{1R}$ has a shorter timelock than $T_{0R}$ to preserve safety. Bob generates a random secret $x$ and sends $H(x)$ to Alice and Carol. Alice then creates a hash-locked transaction $T_2$ that can be redeemed by Carol’s signature and $x$. Carol, in turn, creates $T_3$ that can be redeemed by Bob’s signature and $x$. If Bob redeems $T_3$, he reveals $x$ on-chain, enabling Carol to redeem $T_2$. These \say{claim} transactions act as an enforcement mechanism in case of misbehaviour. 
In the cooperative case, Carol creates $T_4$ to transfer coins to Bob, sends it to him for signing, and publishes it to the blockchain. Once $T_4$ is confirmed, Alice creates $T_5$ to transfer coins to Carol, sends it for her signature, and Carol publishes it. If the protocol stalls, e.g., Bob never claims $T_3$, $T_{0R}$ and $T_{1R}$ are published once their timelocks expire, refunding Alice and Carol.
Thanks to Carol, the transfer of coins between Alice and Bob appears on the blockchain as two independent transfers, given Carol uses two different addresses for the transactions on the two sides. 

\subsection{Stealth Addresses}

Stealth addresses, introduced by van Saberhagen~\cite{van2012stealth}, enhance privacy by preventing observers from linking multiple transactions to the same recipient. Instead of using a static address, the recipient shares a stealth address. For each payment, the sender derives a unique one-time destination address from the sender's ephemeral private key and the recipient’s stealth address via a Diffie--Hellman key exchange. This yields a distinct public key per payment, which only the recipient can identify and spend from by deriving the corresponding private key. Stealth addresses thus break analysis methods that cluster transactions by repeated address reuse.
Privacy-focused cryptocurrencies like Monero use stealth addresses as a core feature~\cite{Noether2015}. In Bitcoin, BIP47 introduces reusable payment codes that allow off-chain generation of stealth addresses~\cite{BIP47}.

\section{Related Work}

Centralised mixing services have been studied extensively. An early study by M{\"{o}}ser, B{\"{o}}hme, and Breuker reported mixed results on their effectiveness~\cite{Moser2013-mn}. Subsequent work 
by Pakki et al.~\cite{PakkiSWBD21} analysed 21 services and identified widespread implementation and security problems. Wu et al.\ model mixing services and propose methods that identify centrally mixed transactions with over 90\% accuracy~\cite{WuHZWLW0021-demystifying}, and later work further improves these detection techniques~\cite{ShojaeinasabMB23}.

Second-generation anonymisation protocols were first studied by Möser and Böhme, who conducted a longitudinal study of their prevalence in 2016~\cite{Moser2017-sb}. 
They surveyed 
Fair exchange, CoinSwap, CoinJoin, and Stealth addresses, using heuristic filtering to analyse their on-chain adoption. 
Their analysis covered the genesis block up to block 418,722 (June 2016).

Möser and Böhme found no evidence of Fair Exchange used in practice, indicating a lack of adoption despite its theoretical feasibility. In contrast, CoinJoin emerged as the most widely used protocol, peaking at around 18 transactions per block (tpb) during 2015--2016. However, its usage declined sharply after 2016, falling to around 7\,tpb, possibly due to changes in wallet behaviour and user priorities. 
CoinSwap, however, did not see much usage until 2016, when it started to be adopted vastly more than CoinJoin, peaking at around 50\,tpb. 

St{\"{u}}tz et al.\ studied the adoption of two popular decentralised CoinJoin implementations, Wasabi and Samourai, within the block range 530k--725k (2018--2022)~\cite{StutzSMHM22}. They found a steady adoption, especially of Samourai after 2020, and showed how address traceability during pre- and post-mixing operations limits the resulting anonymity set sizes. Interestingly, they demonstrate that although the proportion of Wasabi and Samourai transactions stays well below 1\% of total transactions, their absolute value progressively increases. 
Despite its illuminating results, this study only focuses on two specific CoinJoin implementations. 

Möser and Böhme study the JoinMarket brokerage marketplace for organising CoinJoin anonymity pools and discuss the economics of privacy~\cite{MoserB17-price-of-anonymity}. 
Empirical analyses of privacy have also been conducted for other cryptocurrencies, e.g., Zcash~\cite{KapposYMM18-zcash}, and payment networks, e.g., Lightning~\cite{KapposYPKDMM21-lightning}, showing that in practice, anonymity set sizes are usually much smaller than those possible in theory. 
Other works demonstrate how it is often possible to trace 
cross-currency trades~\cite{YousafKM19-tracing}. 

Möser and Böhme's study thus remains the only longitudinal measurement of adoption across multiple second-generation protocols, and ends in 2016. We address this gap by extending the multi-protocol measurement to 2025.

\section{Methodology and Implementation}
To measure the prevalence of privacy-enhancing protocols on the blockchain, we follow a structured pipeline: after downloading the raw blockchain data, we transform it into an efficient data structure to facilitate optimised SQL queries, enabling the application of heuristic filters which parse the blockchain to identify and count transactions possessing the characteristic features of the target protocols. 
To contextualise the resulting time-series prevalence data and investigate the impact of external factors, this timeline is overlaid with significant real-world events. The identification of these events is based on their notoriety and discussion volume within relevant online communities and media platforms.

\subsection{Data Collection}

A local copy of the blockchain is downloaded using the Bitcoin Core client. 
This took 96~hours to download and occupied 780\,GiB of storage. The raw data was parsed using \code{rusty-blockparser}~\cite{Egger2020}, which efficiently extracts and structures the data into four distinct CSV files: blocks, transactions, inputs, and outputs, 
and simplifies the subsequent construction of a relational database. 
Parsing took a further 12~hours on an 8-core Intel i7-7700 machine with 16\,GiB of RAM, and occupied a further 1.3\,TB of storage. 
We then created a DuckDB database, which can be queried through its Python API. To reduce the size of this DuckDB database, unnecessary data was pruned during the creation, as detailed in Table~\ref{tab:CSVPrune}. 

\begin{table}
\caption{Pruning of the Bitcoin Transaction Dataset}
\label{tab:CSVPrune}
\centering
\resizebox{\linewidth}{!}{%
\begin{tabular}{ll}
    \toprule
    \textbf{Table} & \textbf{Columns (those highlighted kept, others pruned out)} \\
    \midrule 
    blocks & \hl{block\_hash}, \hl{height}, version, blocksize, hashPrev, hashMerkleRoot, \hl{nTime}, nBits, nNonce \\
    transactions & \hl{txid}, \hl{hashBlock}, version, \hl{lockTime} \\
    tx\_in & \hl{txid}, \hl{hashPrevOut}, \hl{indexPrevOut}, \hl{scriptSig}, sequence \\
    tx\_out & \hl{txid}, \hl{indexOut}, height, \hl{tValue}, \hl{scriptPubKey}, \hl{address} \\
    \bottomrule
\end{tabular}
}
\end{table}

Protocol transaction occurrences are then counted. Data is aggregated over periods of 1,008 blocks (approximately a week), i.e., half of the Bitcoin difficulty adjustment cycle, a common unit of measurement for longitudinal studies~\cite{Moser2015-sn}, providing a balanced level of granularity, and enabling comparison with prior work. These counts are normalised against total transaction counts in order to account for variations in the number of transactions per block. 
We use the previous results from Möser and Böhme as a sanity check to ensure that our heuristic implementations produce comparable results. 

We analyse the period from block height 430k (September 2016) to 891k (April 2025). 
We call a protocol \emph{detectable} if its transactions can be identified from public on-chain data alone. As many modern privacy techniques (e.g., Taproot-based or off-chain ones) are not detectable, we focus on CoinJoin, CoinSwap, CoinShuffle, and Stealth Addresses.

We aimed for a reproducible measurement approach that was consistent over a multi-year timeline. Proprietary third-party services, e.g., Chainalysis and Elliptic, although powerful for targeted investigations, are ill-suited for this purpose. They incorporate off-chain data, and use models which are susceptible to changes, introducing inconsistency for longitudinal study. We therefore chose to adopt an on-chain, heuristic-based method,  aligning broadly with Möser and Böhme's~\cite{Moser2017-sb}, ensuring reproducibility and comparability with previous results. 

\emph{CoinJoin.}
Because transactions must have a minimum of two participants, and ignoring the rare edge case where a party has an exact UTXO quantity to spend, we expect a CoinJoin transaction to have at least four outputs: two for spending, and two as change. In general, the number of inputs must be at least half the number of outputs. 
As this broad heuristic may also match other transactions with many outputs (e.g., batched payments or self-transfers), the detected volumes should be interpreted as an upper bound on potential CoinJoins.

\emph{CoinSwap.}
Each CoinSwap requires two 2-of-2 multi-signature transactions, so we select multisig outputs (identified by \opcheckmultisig\ or its hex representation), discarding transactions with more than two such outputs. As the protocol must complete before the locktime of the refund transactions elapses, we require these outputs to be spent within 30 blocks of their creation. We then pair each output with another such output created within the next 30 blocks, restricting their values: a payment from Carol (to Bob) should be slightly less than the payment from Alice (to Carol) due to transaction fees. Möser and Böhme proposed a leniency of 0.002\,\btc. We used a stricter 0.0002\,\btc, as the rise in \btc\ value means fees correspond to smaller \btc\ amounts. We do not enforce that the two parties are unconnected in the transaction graph.

\emph{CoinShuffle.}
In a CoinShuffle transaction, we need at least 3 distinct participants to avoid trivial internal de-anonymisation; the number of outputs must exactly match the number of inputs; and all outputs must be almost exactly the same value. These restrictions are common between the standard CoinShuffle protocol and its variations (e.g., \textit{CoinShuffle++}~\cite{Ruffing2016-lx}, \textit{ValueShuffle}~\cite{Ruffing2017-pf} and WabiSabi~\cite{ficsor2021wabisabi}), hence our heuristic technically detects equal-output, CoinJoin-like transactions. 

\emph{Stealth Addresses.}
Möser and Böhme detected stealth addresses by simply detecting raw public keys in \opreturn\ statements. This introduced unexpected challenges, requiring three refinement iterations. 
The main challenge stemmed from the lack of a standardised format for embedding public keys in \opreturn\ fields. As a result, implementations can vary significantly, making detection
inherently susceptible to false positives
(e.g., hex-encoded text or images, may contain byte sequences resembling compressed public keys). The refinement therefore adjusted the specificity of this premise. 

\emph{The first heuristic} targeted only outputs containing the exact \opreturn\ lengths, e.g., 70 and 138 bytes, and public key repetition patterns that were theorised to be associated with stealth address protocols, e.g., repeated public key prefixes e.g., \code{\%02020202\%}, which were used by early implementations as a way of easily identifying their transactions.
This yielded only about 5,000 matches, too few for time-series analysis. 

\emph{The second heuristic} was developed 
by relaxing constraints on script length and content, and was designed to identify outputs containing non-standard data, a characteristic of stealth address protocols.
This produced many detections, but manual inspection suggested many false positives. 

\emph{The third heuristic} was designed to balance precision and recall. This version moves from general pattern matching to targeting the exact byte-level structures of specific, known privacy protocols that are functionally identical to Stealth Addresses. The primary target was the BIP47 Reusable Payment Code standard~\cite{BIP47}. While not a pure stealth address protocol, BIP47 operates under a similar premise, enabling a payer to generate a unique, one-time address for a payee without the payee revealing their main public address in advance. Critically, BIP47 mandates a specific on-chain footprint which can be traced accurately. 
Hence, this heuristic targeted specific opcodes, byte prefixes, and fixed lengths. The results presented in Section~\ref{sec:results} are for the second and third heuristics. 

These heuristics are summarised in Table~\ref{tab:Heuristics} and implemented in SQL. A full code repository is available at 
\url{https://github.com/bhawks3/BTC-Privacy-Analysis}.

\begin{table}
  \centering
  \small 
  \caption{Summary of Criteria for Protocol Identification}
  \label{tab:Heuristics}
  \resizebox{\linewidth}{!}{%
  \begin{tabular}{l@{\quad}p{1\columnwidth}}
    \toprule
    \textbf{Protocol} & \textbf{Criteria} \\
    \midrule
    CoinJoin & 
        1)~A minimum of 4 outputs (at least two participants, each with a payment and a change output), and 
        2)~Inputs at least half the number of outputs.
    \\
    \midrule 
    CoinSwap & 
        1)~Multisig outputs, at most two per transaction, 
        2)~Spent within 30 blocks of creation (before locktime expiration), and 
        3)~Paired with another such output created within the next 30 blocks, with values within 0.0002\,\btc.
    \\
    \midrule
    CoinShuffle & 
        1)~At least 3 participants, 
        2)~Inputs and outputs must match exactly, and 
        3)~All outputs must be close to the same value. 
    \\ 
    \midrule
    Stealth Address & 1)~BIP47 structure in \opreturn\ field (Heuristic 3; Heuristics 1--2 are described in the text).
    \\
    \bottomrule
  \end{tabular}
  }
\end{table}

\subsection{Event Gathering}
\begin{figure*}[t]
    \centering
    \begin{minipage}{\columnwidth}
        \centering
        \includegraphics[width=\linewidth]{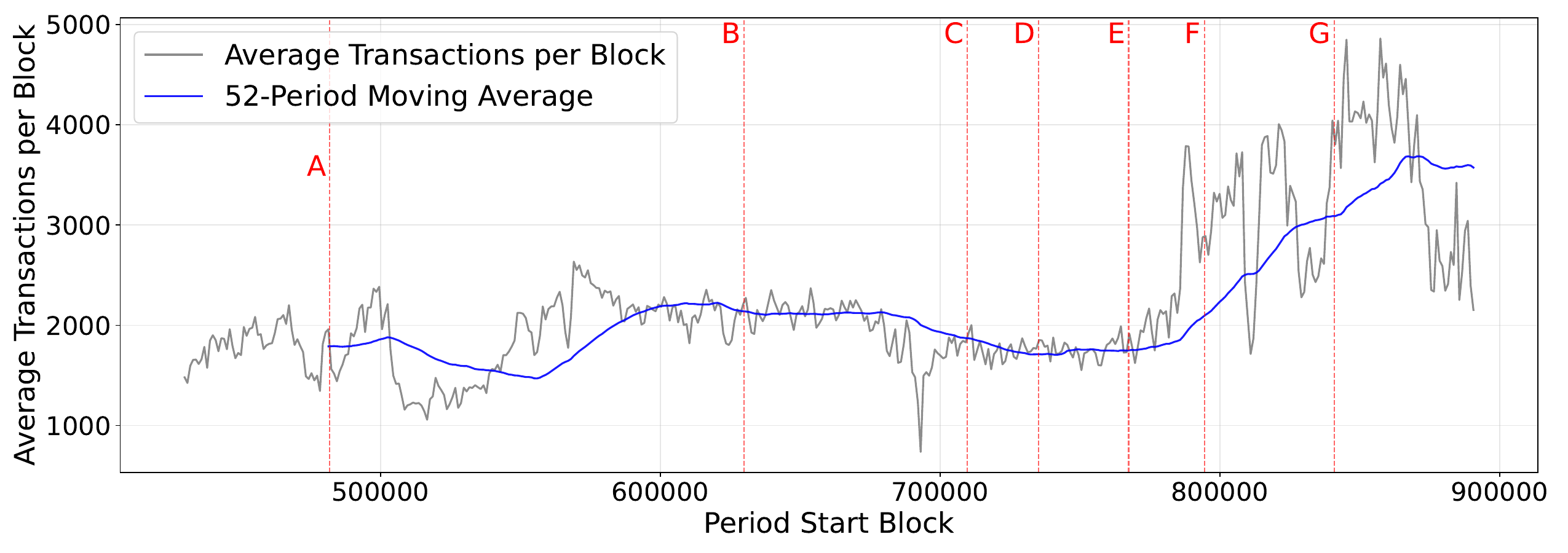}
        \caption{BTC Transactions per Block over 1,008 block Periods}
        \label{fig:Blockstats}
    \end{minipage}%
    \hfill
    \begin{minipage}{\columnwidth}
        \centering
        \includegraphics[width=\linewidth]{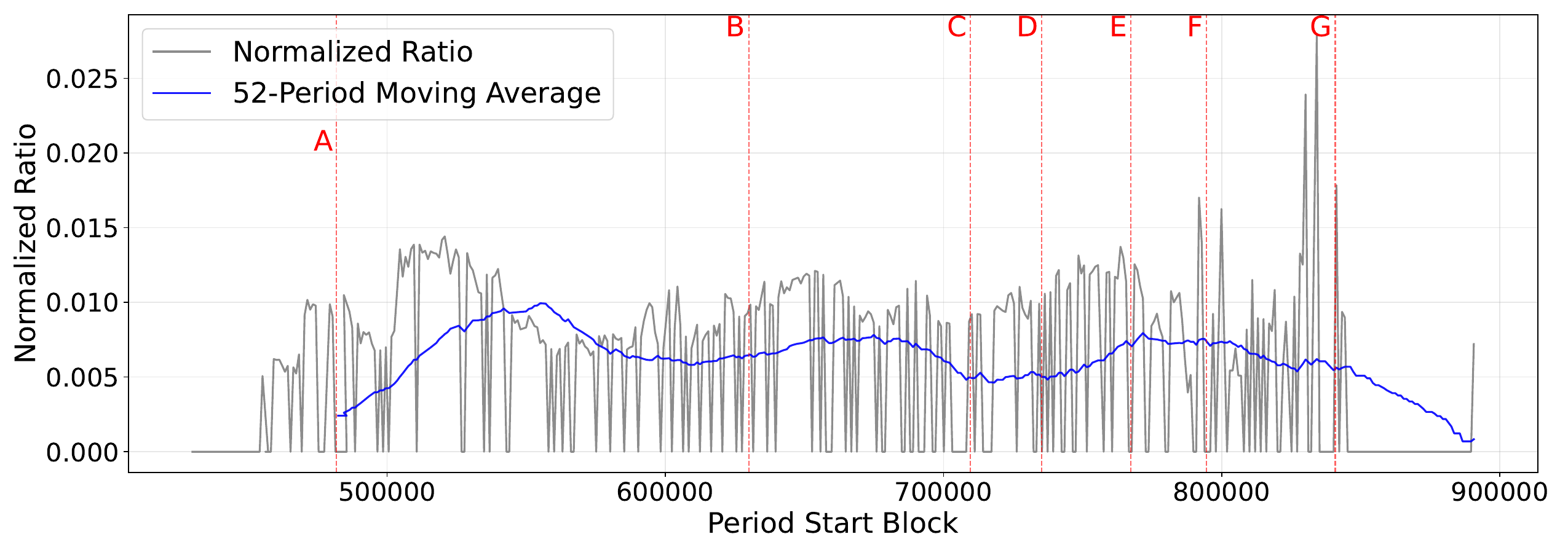}
        \caption{Normalised CoinJoin Transactions (Fraction of All Transactions)}
        \label{fig:NormCoinJoin}
    \end{minipage}
\end{figure*}

The events to be considered for analysis were identified through a qualitative evaluation of historical significance and community impact. We began by consulting established relevant historical timelines, such as the price history and event chronicles provided by Bitcoin Magazine~\cite{Mulcahy2023-ic}. This provided an initial shortlist of major market, regulatory, and technical milestones.
Each potential event from this list was then qualitatively researched across key online forums, including \code{r/CryptoCurrency}, \code{r/Bitcoin}, \code{r/Privacy}, and \code{bitcointalk.com}. 
Events were selected based on evidence of significant controversy, polarising debate, or being widely perceived as a threat to either Bitcoin value, or privacy in the network  (and therefore most likely to have influenced user behaviour).

\section{Results and Analysis}
\label{sec:results}
The events deemed to have potential for significantly impacting protocol adoption were identified as follows: 
\begin{itemize}
\item[A:] 481,824 (Aug 24, 2017): SegWit is deployed~\cite{bip141-segwit}.
\item[B:] 630,000 (May 11, 2020): Block Reward decreased (12.5\,\btc\ to 6.25\,\btc)~\cite{halving-chainalysis}.
\item[C:] 709,632 (Nov 14, 2021): Taproot upgrade activated~\cite{Unknown2021-mj}.
\item[D:] 735,300 (May 7, 2022): Collapse of Terra/Luna ecosystem begins~\cite{terra-bitstamp}.
\item[E:] 767,430 (Dec 14, 2022): First Ordinals inscription~\cite{UkuriaOC2023-ck}.
\item[F:] 794,500 (Jun 15, 2023): BlackRock files for Bitcoin trust~\cite{Helms2023-cg}.
\item[G:] 840,850 (Apr 25, 2024): FBI warns against using unregistered cryptocurrency money-transmitting services~\cite{ic3-fbi}. 
\end{itemize}

Upgrades such as SegWit (Event A) and Taproot (Event C) likely improved Bitcoin's privacy and efficiency, indirectly increasing interest in privacy solutions. The halving (Event B) and market collapses such as Terra/Luna (Event D) tend to trigger speculation and privacy concerns, potentially increasing privacy tool adoption. Institutional involvement such as BlackRock's ETF filing (Event F) bolsters market confidence, drawing in more users, some of whom may adopt privacy-enhancing solutions. The rise of Ordinals (Event E) increased competition for block space and fees, potentially making privacy techniques more costly. Regulatory scrutiny such as the FBI warning (Event G) may push users towards more robust, possibly less detectable, privacy tools. While we hypothesise that these events affected protocol adoption, temporal proximity alone establishes correlation, not causation.

Figure~\ref{fig:Blockstats} shows the total Bitcoin transactions per block throughout the studied period, 
reflecting the overall demand for and usage of Bitcoin.
From around block 780k, there is a significant increase in the average number of transactions per block. This surge is most likely attributable to the \textit{Ordinals phenomenon} (Event E), a period of rapid adoption of Bitcoin-based digital artifacts. The timing aligns well with the peak of market activity surrounding these innovations, which began with the first inscription at block 767,430. 

Ordinals operate by inscribing data directly onto individual satoshis within transactions. This process generates a high volume of data-heavy transactions that compete for block space, potentially leading to the observed congestion and increase in the number of transactions per block. Crucially, this was coupled with a significant spike in transaction fees, a key indicator of intense demand for block inclusion that further corroborates this finding~\cite{btcavgtx}.

The Ordinals phenomenon represents a distinct shift in Bitcoin usage, from peer-to-peer payments towards data inscription and digital collectibles. This seems to have had a dual effect on our analysis: 1)~\emph{economic crowding out}, as higher fees may make privacy techniques prohibitively expensive, particularly those requiring multiple transactions (e.g., CoinSwap) or large multi-input transactions (e.g., CoinJoin); and 2)~\emph{anonymity set noise}, as the influx of non-financial transactions could provide cover for privacy-seeking users but also complicates on-chain analysis.

\subsection{CoinJoin}
The heuristic analysis identified a total of 5,048,184 CoinJoin transactions across the full study period. As Figure~\ref{fig:NormCoinJoin} shows, the relative adoption of CoinJoin remained exceptionally low throughout, consistently fluctuating between 0\% and 0.8\% of total Bitcoin transactions, in line with Möser and Böhme's results~\cite{Moser2017-sb} despite our broader heuristic. This persistent sub-1\% rate suggests that CoinJoin has remained a niche solution, never achieving widespread adoption.

Usage fluctuates between consistent periods, e.g., after the SegWit lock-in (Event A), and erratic phases, particularly before the Terra/Luna collapse (Event D) and after the BlackRock ETF filing (Event F). The largest spikes in activity occurred during volatile periods; after Event F, detectable usage dropped to zero before surging to isolated peaks of 2.5\% of network transactions. These sharp fluctuations indicate CoinJoin adoption is episodic and reactive. 

There is a noticeable spike at the 830k block range (February--March 2024), which may indicate a reaction to two developments: intensifying regulatory pressure on privacy tools, notably a US proposal to designate cryptocurrency mixing a primary money laundering concern (October 2023), and the Spot Bitcoin ETF approval (January 2024) that followed Event F. Rather than suppressing activity immediately, the regulatory pressure likely induced a short-term \say{last call} effect among existing users, incentivising them to make a final use of privacy-preserving techniques.

The general trend (represented by the 52-period moving average), suggests a steady decline in usage. This trend was cemented in 2024, appearing to coincide with the FBI warnings (Event G), where it can be seen that the detected usage rapidly falls off. Event G closely followed the April 2024 halving (block 840,000), which also launched the Runes token protocol and briefly pushed fees to record levels, likely compounding this decline. In fact, our heuristic detects almost no CoinJoin activity from shortly after Event G onwards. 

\subsection{CoinSwap}
A total of 23,321,415 CoinSwap transactions were detected across the study period, which is significantly higher than CoinJoin. However, this higher transaction count does not directly imply higher user adoption. CoinJoin can aggregate many participants into a single transaction, whereas each CoinSwap detection represents a paired exchange involving exactly three parties. Therefore, if the same number of individuals were using both protocols, we would expect far more CoinSwap transactions, than CoinJoin transactions. 

As a rough illustration, assuming around 50 participants per CoinJoin~\cite{PulpCattel_Wasabi_Observatory} and 3 per CoinSwap, CoinJoin would account for over three times more participations, consistent with lower real-world adoption of the more complex CoinSwap.

\begin{figure}[t]
    \centering
    \includegraphics[width=\columnwidth]{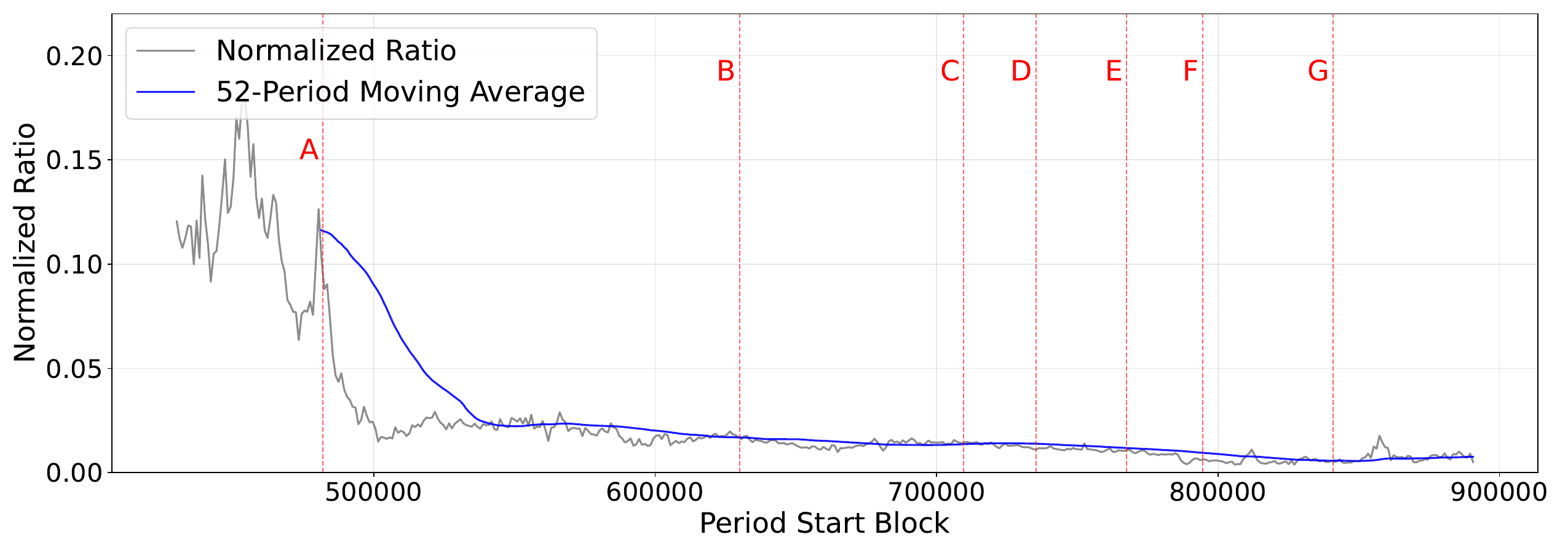}
    \caption{Normalised CoinSwap Transactions (Fraction of All Transactions)}
    \label{fig:NormCoinSwap}
\end{figure}

In Figure~\ref{fig:NormCoinSwap}, we see an initial adoption rate that peaks at 17\% of all transactions (or around 351\,tpb), higher than the upper bounds proposed by Möser and Böhme. This higher number of detections may be explained by subtle differences in heuristic implementation (mainly we do not enforce full unconnectedness in the transaction tree). Eventually the detections settled down to reflect amounts similar to those suggested by Möser and Böhme, fluctuating from 20--30\,tpb.

After SegWit activation (Event A), detections dropped drastically. The SegWit upgrade moves 2-of-2 multisig unlocking conditions from \code{scriptPubKey}/\code{scriptSig}, to a separate witness field which we do not parse. Our heuristic, which relied on detecting \opcheckmultisig\ (or its hex representation, \code{ae}) in those fields, therefore fails to detect native SegWit CoinSwaps.

As SegWit is optional, residual detections may stem from the few remaining non-SegWit users; however, the small anonymity set this implies makes them more likely to be non-privacy uses of 2-of-2 multisig (e.g., coloured coins, excludable by removing transactions with \opreturn\ outputs~\cite{Moser2017-sb}).

\subsection{CoinShuffle}
\begin{figure}[t]
    \centering
    \includegraphics[width=\columnwidth]{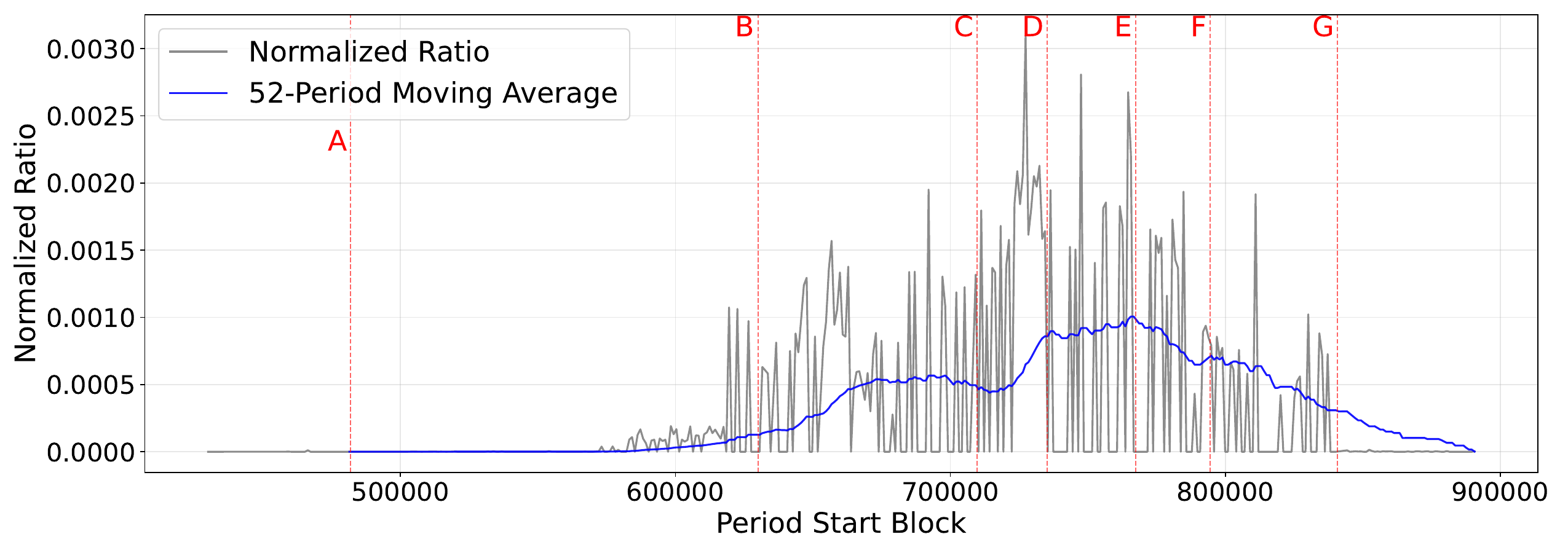}
    \caption{Normalised CoinShuffle Transactions (Fraction of All Transactions)}
    \label{fig:NormCoinShuffle}
\end{figure}

The CoinShuffle heuristic analysis results, shown in Figure~\ref{fig:NormCoinShuffle}, reveal an unexpected finding: the detected transaction volume mirrors the
operational history of coordinator-based CoinJoin services such as Wasabi Wallet and Samourai Whirlpool, whose equal-output transactions have an on-chain structure functionally identical to that of CoinShuffle. 

Although CoinShuffle was proposed in 2014, sustained activity only emerged around block 583k (mid-2019), coinciding with the public launch of Samourai Whirlpool (June 2019), and rose sharply in early 2020 when Whirlpool became available on Samourai's mobile wallet. This adoption grew steadily, reaching its peak at around block 730k (early 2022), before a pronounced decline commenced. 
This decline correlates with a period of intense external pressure on cryptocurrency privacy tools. The peak coincides with Wasabi's coordinator starting to refuse coins flagged as illicit (March 2022), and the decline continued after the Terra/Luna collapse (Event D), which sparked an increase in regulatory scrutiny across the sector, and the US sanctioning of the Ethereum mixer Tornado Cash (August 2022). In addition, Wasabi's switch to the WabiSabi protocol (June 2022), which allows varied output amounts, likely moved its rounds outside our equal-output heuristic. The trend concludes with an abrupt termination of all detectable activity coinciding with the shutdown of Samourai Whirlpool following the arrest of its developers in April 2024, the FBI warning (Event G), and the end of Wasabi's coordination service shortly after. This clear endpoint suggests regulatory actions can affect even non-custodial privacy tools.

Moreover, the analysis reveals that several high-profile events in the broader Bitcoin timeline appear to have had a negligible correlative effect on the adoption of CoinShuffle-style transactions. 
This decoupling suggests privacy tool adoption is possibly shaped more by operational viability and legal standing than by general market cycles or protocol upgrades.

\subsection{Stealth Addresses}
Three progressively refined heuristics yielded distinctly different outcomes, collectively underlining the challenge in detecting stealth address transactions. 

A brief analysis of the first heuristic's output revealed that its 5k detections were not spread uniformly over time. The transactions were clustered into three distinct, short-lived peaks, each lasting approximately one week and occurring at seemingly random intervals. The absence of a sustained, low-level baseline of activity indicates that these detections likely represent anomalous events, e.g., the short-term testing of a protocol or, more likely, non-stealth data embedding, rather than evidence of stealth address usage.

\begin{figure}[t]
    \centering
    \includegraphics[width=\columnwidth]{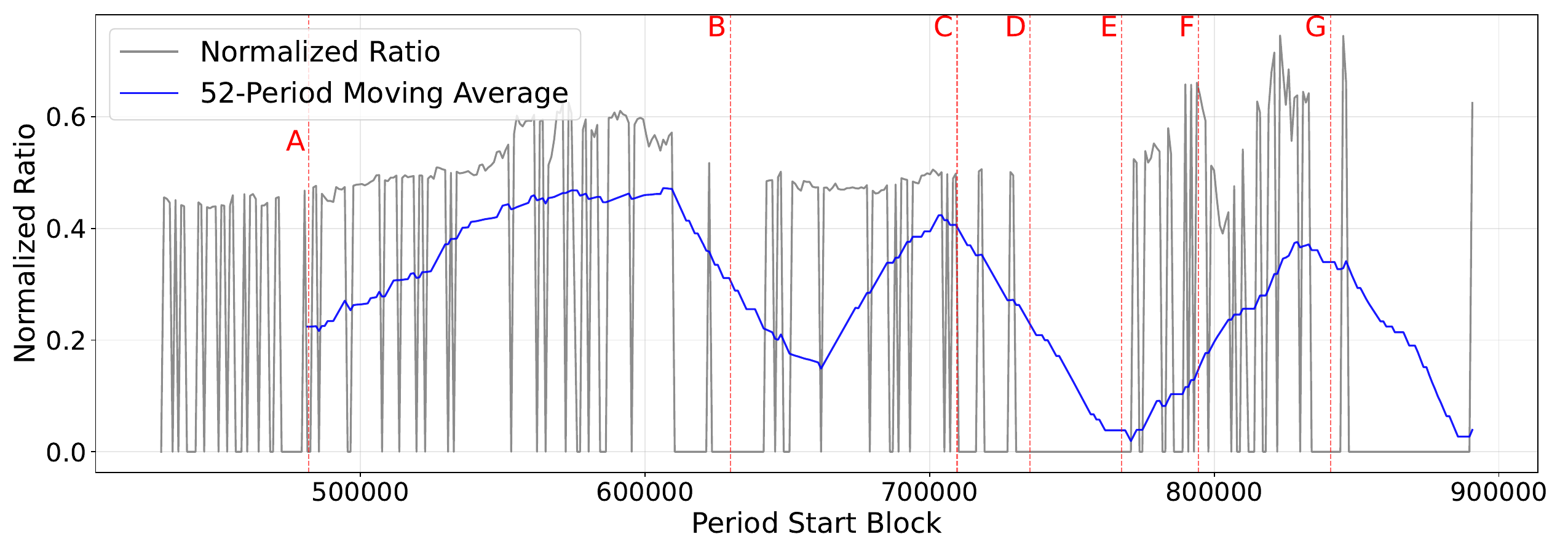}
    \caption{Normalised Stealth Address Transactions (Fraction of All Transactions; Heuristic 2)}
    \label{fig:NormStealth}
\end{figure}

The second heuristic produced many detections (Fig.~\ref{fig:NormStealth}), but spikes correlated strongly with external events unrelated to privacy (e.g., Ordinals protocol). This, combined with low adoption reported by Möser and Böhme, suggests false positives: the heuristic's generic patterns (\code{\%02\%}, \code{\%03\%}) matched other embedded data, e.g., from data-embedding protocols that surged alongside Ordinals, not stealth address metadata.

In contrast to the noisy output of the second heuristic, the application of the third heuristic targeting BIP47 transactions returned a null result: no transactions were detected across the entire studied block range. This could be indicative of two possibilities:
\begin{inparaenum}[1)]
\item \textit{Genuine absence}: The BIP47 protocol, and any other protocol sharing its exact standardised on-chain footprint, have seen no adoption on the blockchain over this study period.
\item \textit{Evolution beyond detection}: Implementations may have evolved to use a different, non-standard method for embedding payment codes, outside this heuristic's strict parameters. Newer standards such as Silent Payments (BIP352, 2023) leave no on-chain footprint for such heuristics. 
\end{inparaenum}

That the first heuristic detected 5k transactions while the more precise third heuristic found none is not a contradiction: the first heuristic's length and byte-repetition filters also match arbitrary non-stealth data, whereas the third targets the exact BIP47 specification. Its null result therefore suggests a genuine absence of BIP47 usage, although we cannot rule out that some BIP47 transactions escape our heuristic.

Although BIP47 offers a standard for stealth-like payments, we find no detectable uptake. This is consistent with claims in the literature that the lack of a widely adopted standard has been a critical barrier to the adoption of privacy-enhancing technologies~\cite{Meiklejohn2013-lg,Moser2017-sb}.
However, it is not possible to definitively conclude an abandonment of stealth addresses over an evolution beyond detectable means, as there remains potential for non-standardised implementations of the protocol, which our heuristics have not captured.

\subsection{Protocol Competition}
\begin{figure}[t]
    \centering
    \includegraphics[width=\columnwidth]{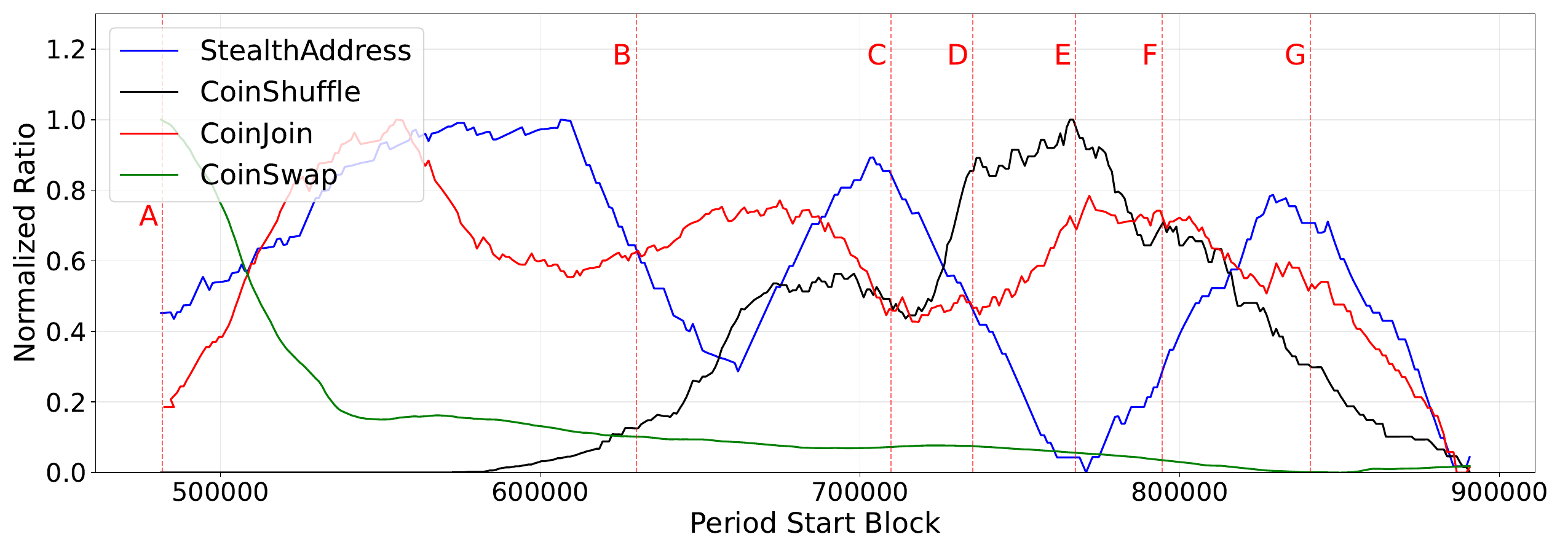}
    \caption{52-Period Moving Averages of Normalised Transactions, Each Scaled to Its Maximum}
    \label{fig:ComparisonAllNorm}
\end{figure}

Figure~\ref{fig:ComparisonAllNorm} shows how each protocol evolves over time relative to the others. User preferences remain roughly stable, indicating that abandoning one protocol does not typically lead to adopting another. Each protocol thus appears to occupy a relatively stable niche in the Bitcoin ecosystem (which is important for building anonymity sets) rather than competing for dominance. The stability of these proportions suggests users adopt privacy techniques for specific use cases, not by opportunistically switching between them. Consequently, privacy mechanisms seem driven more by user intent or technical requirements than by changes in available alternatives.

However, the lack of observable substitution between protocols does not preclude external factors influencing adoption. Regulatory and protocol changes (e.g., Events A, C and G) or improvements in blockchain analysis techniques may suppress detectable activity across all privacy-enhancing methods simultaneously. Conversely, advances in wallet software or protocol upgrades could reinforce the usage of a single technique without affecting others. 

Trends of popularity between protocols have remained similar to those discussed by Möser and Böhme. CoinJoin is by far the preferred technique, most likely due to being one of the first trustless protocols and low technical overhead. 
Protocol utilisation remains low as non-private activity (e.g., exchange settlements and Ordinals) dilutes usage ratios and advanced users move to undetectable methods such as Taproot-based protocols. Fluctuations likely coincide with periods of low fees, when privacy is cheaper and these users more visible.

\subsection{Limitations}
Due to the lack of available ground-truth data, we have been unable to determine accuracy measures of our heuristics (e.g., false positive/negative rates), so our counts indicate trends rather than exact volumes. False positives are most likely for the broad CoinJoin heuristic (e.g., batched payments or consolidations), for CoinSwap (non-privacy uses of 2-of-2 multisig), and for the second Stealth Address heuristic (arbitrary embedded data). False negatives arise as we do not parse witness data, missing SegWit- and Taproot-based transactions, and from implementations deviating from the assumed structures (e.g., CoinJoins with fewer than four outputs or non-standard Stealth Address encodings).
The simultaneous disappearance of detections across heuristics around 2024 may also partly reflect a shift to transaction formats our heuristics do not capture.
The observed alignment between Events A--G and fluctuations in detection volumes suggests a potential relationship, but not causality. Moreover, our event list is not exhaustive, so other events may also have influenced adoption.

\section{Conclusion}
Overall, adoption of privacy-preserving techniques in the Bitcoin network remained low. CoinJoin and CoinShuffle saw steady use until 2024, when regulatory changes appear to have pushed privacy activity beyond detectable methods. Fee spikes (Ordinals, Runes) and protocol changes (WabiSabi, Silent Payments) likely contributed too. It is unclear whether this reflects a shift to stronger, less detectable Bitcoin privacy protocols~\cite{glaeser2022foundations},~\cite{Heilman2017-nk} or privacy-focused coins, e.g., Monero or Zcash, or a drop in user demand for privacy.

Future work could examine whether the detected drop in adoption reflects a genuine decline in user demand for privacy or a migration to more advanced, less detectable methods. Validating our heuristics against ground-truth data, e.g., known Whirlpool or Wasabi transactions, would quantify their false positive and negative rates. Analysing usage trends in alternative networks, such as Monero or Zcash, alongside Bitcoin could clarify how they have evolved in parallel and help confirm or reject the migration hypothesis. Extending monitoring to dual-chain analysis would also shed light on privacy in newer ecosystems such as Lightning and Taproot.

\section*{Acknowledgements}
The Viking cluster, a high performance compute facility provided by the University of York, was used during this project. We are grateful for computational support from the University of York IT Services and the Research IT team. 

\bibliographystyle{IEEEtran}
\bibliography{references}

\begin{thebibliography}{10}
\providecommand{\url}[1]{#1}
\csname url@samestyle\endcsname
\providecommand{\newblock}{\relax}
\providecommand{\bibinfo}[2]{#2}
\providecommand{\BIBentrySTDinterwordspacing}{\spaceskip=0pt\relax}
\providecommand{\BIBentryALTinterwordstretchfactor}{4}
\providecommand{\BIBentryALTinterwordspacing}{\spaceskip=\fontdimen2\font plus
\BIBentryALTinterwordstretchfactor\fontdimen3\font minus
  \fontdimen4\font\relax}
\providecommand{\BIBforeignlanguage}[2]{{%
\expandafter\ifx\csname l@#1\endcsname\relax
\typeout{** WARNING: IEEEtran.bst: No hyphenation pattern has been}%
\typeout{** loaded for the language `#1'. Using the pattern for}%
\typeout{** the default language instead.}%
\else
\language=\csname l@#1\endcsname
\fi
#2}}
\providecommand{\BIBdecl}{\relax}
\BIBdecl

\bibitem{nakamoto2008bitcoin}
S.~Nakamoto, ``Bitcoin: A peer-to-peer electronic cash system,'' 2008,
  available online from \href{www.bitcoin.org/en/bitcoin-paper}{bitcoin.org}.

\bibitem{ReidH11}
F.~Reid and M.~Harrigan, ``An analysis of anonymity in the {Bitcoin} system,''
  in \emph{SocialCom/PASSAT}.\hskip 1em plus 0.5em minus 0.4em\relax {IEEE}
  ComSoc, 2011, pp. 1318--1326.

\bibitem{Ron2013-ez}
D.~Ron and A.~Shamir, ``Quantitative analysis of the full {Bitcoin} transaction
  graph,'' in \emph{FC'13}, ser. LNCS, vol. 7859.\hskip 1em plus 0.5em minus
  0.4em\relax Springer, 2013, pp. 6--24.

\bibitem{Meiklejohn2013-lg}
S.~Meiklejohn, M.~Pomarole, G.~Jordan, K.~Levchenko, D.~McCoy, G.~M. Voelker,
  and S.~Savage, ``A fistful of {Bitcoins}: Characterizing payments among men
  with no names,'' in \emph{ACM IMC'13}, 2013, pp. 127--140.

\bibitem{Harrigan2016-ak}
M.~Harrigan and C.~Fretter, ``The unreasonable effectiveness of address
  clustering,'' in \emph{UIC/ATC/ScalCom/CBDCom/IoP/SmartWorld}.\hskip 1em plus
  0.5em minus 0.4em\relax {IEEE}, 2016, pp. 368--373.

\bibitem{Moser2017-sb}
M.~M{\"{o}}ser and R.~B{\"{o}}hme, ``Anonymous alone? measuring {Bitcoin}'s
  second-generation anonymization techniques,'' in \emph{EuroS{\&}P
  Workshops}.\hskip 1em plus 0.5em minus 0.4em\relax {IEEE}, 2017, pp. 32--41.

\bibitem{Maxwell2013coinjoin}
G.~Maxwell, ``Coinjoin: Bitcoin privacy for the real world,'' Bitcoin Forum,
  2013, \url{https://bitcointalk.org/index.php?topic=279249.0}.

\bibitem{Ruffing2014-em}
T.~Ruffing, P.~Moreno{-}Sanchez, and A.~Kate, ``{CoinShuffle}: Practical
  decentralized coin mixing for {Bitcoin},'' in \emph{{ESORICS}'14}, ser. LNCS,
  vol. 8713.\hskip 1em plus 0.5em minus 0.4em\relax Springer, 2014, pp.
  345--364.

\bibitem{Corrigan2010-ed}
H.~Corrigan{-}Gibbs and B.~Ford, ``Dissent: accountable anonymous group
  messaging,'' in \emph{{CCS}}.\hskip 1em plus 0.5em minus 0.4em\relax {ACM},
  2010, pp. 340--350.

\bibitem{Barber2012-yx}
S.~Barber, X.~Boyen, E.~Shi, and E.~Uzun, ``Bitter to better -- how to make
  {Bitcoin} a better currency,'' in \emph{FC'12}, ser. LNCS, vol. 7397.\hskip
  1em plus 0.5em minus 0.4em\relax Springer, 2012, pp. 399--414.

\bibitem{maxwell-coinswap}
G.~Maxwell, ``{CoinSwap}: Transaction graph disjoint trustless trading,''
  Bitcoin Forum, 2013, \url{https://bitcointalk.org/index.php?topic=321228}.

\bibitem{van2012stealth}
N.~van Saberhagen, ``Cryptonote v 2.0,'' Whitepaper, 2013, available online
  from
  \href{https://web.getmonero.org/resources/research-lab/pubs/cryptonote-whitepaper.pdf}{getmonero.org}.

\bibitem{Noether2015}
S.~Noether, ``Ring confidential transactions,'' {IACR} Cryptology {ePrint}
  Archive, Paper 2015/1098, 2015, \url{https://ia.cr/2015/1098}.

\bibitem{BIP47}
J.~Ranvier, ``{BIP 47}: Reusable payment codes for hierarchical deterministic
  wallets,'' 2015, \url{https://bips.dev/47}.

\bibitem{Moser2013-mn}
M.~M{\"{o}}ser, R.~B{\"{o}}hme, and D.~Breuker, ``An inquiry into money
  laundering tools in the {Bitcoin} ecosystem,'' in \emph{eCrime}.\hskip 1em
  plus 0.5em minus 0.4em\relax {IEEE}, 2013, pp. 1--14.

\bibitem{PakkiSWBD21}
J.~Pakki, Y.~Shoshitaishvili, R.~Wang, T.~Bao, and A.~Doup{\'{e}}, ``Everything
  you ever wanted to know about {Bitcoin} mixers (but were afraid to ask),'' in
  \emph{FC'21, LNCS 12674}.\hskip 1em plus 0.5em minus 0.4em\relax Springer,
  2021, pp. 117--146.

\bibitem{WuHZWLW0021-demystifying}
L.~Wu, Y.~Hu, Y.~Zhou, H.~Wang, X.~Luo, Z.~Wang, F.~Zhang, and K.~Ren,
  ``Towards understanding and demystifying {Bitcoin} mixing services,'' in
  \emph{{WWW}}.\hskip 1em plus 0.5em minus 0.4em\relax {ACM}\,/\,{IW3C2}, 2021,
  pp. 33--44.

\bibitem{ShojaeinasabMB23}
A.~Shojaeinasab, A.~P. Motamed, and B.~Bahrak, ``Mixing detection on {Bitcoin}
  transactions using statistical patterns,'' \emph{{IET} Blockchain}, vol.~3,
  no.~3, pp. 136--148, 2023.

\bibitem{StutzSMHM22}
R.~St{\"{u}}tz, J.~Stockinger, P.~Moreno{-}Sanchez, B.~Haslhofer, and
  M.~Maffei, ``Adoption and actual privacy of decentralized {CoinJoin}
  implementations in {Bitcoin},'' in \emph{AFT'22}.\hskip 1em plus 0.5em minus
  0.4em\relax {ACM}, 2022, pp. 254--267.

\bibitem{MoserB17-price-of-anonymity}
M.~M{\"{o}}ser and R.~B{\"{o}}hme, ``The price of anonymity: empirical evidence
  from a market for {Bitcoin} anonymization,'' \emph{J. Cybersecur.}, vol.~3,
  no.~2, pp. 127--135, 2017.

\bibitem{KapposYMM18-zcash}
G.~Kappos, H.~Yousaf, M.~Maller, and S.~Meiklejohn, ``An empirical analysis of
  anonymity in {Zcash},'' in \emph{{USENIX} Security'18}, 2018, pp. 463--477.

\bibitem{KapposYPKDMM21-lightning}
G.~Kappos, H.~Yousaf, A.~M. Piotrowska, S.~Kanjalkar, S.~Delgado{-}Segura,
  A.~Miller, and S.~Meiklejohn, ``An empirical analysis of privacy in the
  {Lightning Network},'' in \emph{FC'21}, ser. LNCS, vol. 12674.\hskip 1em plus
  0.5em minus 0.4em\relax Springer, 2021, pp. 167--186.

\bibitem{YousafKM19-tracing}
H.~Yousaf, G.~Kappos, and S.~Meiklejohn, ``Tracing transactions across
  cryptocurrency ledgers,'' in \emph{{USENIX} Security'19}, 2019, pp. 837--850.

\bibitem{Egger2020}
M.~Egger, ``Rusty-blockparser,''
  \url{https://crates.io/crates/rusty-blockparser}.

\bibitem{Moser2015-sn}
M.~M{\"{o}}ser and R.~B{\"{o}}hme, ``Trends, tips, tolls: {A} longitudinal
  study of bitcoin transaction fees,'' in \emph{FC'15 Workshops}, ser. LNCS,
  vol. 8976.\hskip 1em plus 0.5em minus 0.4em\relax Springer, 2015, pp. 19--33.

\bibitem{Ruffing2016-lx}
T.~Ruffing, P.~Moreno-Sanchez, and A.~Kate, ``{P2P} mixing and unlinkable
  bitcoin transactions,'' {IACR} {ePrint} 2016/824, 2016,
  \url{https://ia.cr/2016/824}.

\bibitem{Ruffing2017-pf}
T.~Ruffing and P.~Moreno{-}Sanchez, ``{ValueShuffle}: Mixing confidential
  transactions for comprehensive transaction privacy in {Bitcoin},'' in
  \emph{FC'17 Workshops, LNCS~10323}.\hskip 1em plus 0.5em minus 0.4em\relax
  Springer, 2017, pp. 133--154.

\bibitem{ficsor2021wabisabi}
{\' A}.~Fics{\' o}r, I.~A. Seres, Y.~Kogman, and L.~Ontivero, ``{WabiSabi}:
  Centrally {Coordinated} {CoinJoins} with {Variable} {Amounts},''
  \emph{Cryptoeconomic Systems}, vol.~1, no.~2, 2021.

\bibitem{Mulcahy2023-ic}
C.~Mulcahy, ``Bitcoin price history: 2009--2025,'' Bitcoin Magazine, available
  online from
  \href{https://bitcoinmagazine.com/guides/bitcoin-price-history}{bitcoinmagazine.com}.

\bibitem{bip141-segwit}
E.~Lombrozo, J.~Lau, and P.~Wuille, ``{BIP 141}: Segregated witness (consensus
  layer),'' 2015, \url{https://bips.dev/141}.

\bibitem{halving-chainalysis}
{Chainalysis Team}, ``What you need to know about the {Bitcoin} halving,''
  2024, available online from
  \href{https://www.chainalysis.com/blog/bitcoin-halving-2024}{chainalysis.com}.

\bibitem{Unknown2021-mj}
------, ``Bitcoin’s {Taproot} upgrade: Everything you need to know,'' 2021,
  available online from
  \href{https://www.chainalysis.com/blog/bitcoin-taproot-upgrade}{chainalysis.com}.

\bibitem{terra-bitstamp}
Bitstamp, ``Terra network collapse,'' 2024, available online from
  \href{https://www.bitstamp.net/en-gb/learn/crypto-101/terra-network-collapse}{bitstamp.net}.

\bibitem{UkuriaOC2023-ck}
{UkuriaOC}, Checkmate, and Glassnode, ``Ordinal theory and the rise of bitcoin
  inscriptions,'' 2023, available online from
  \href{https://insights.glassnode.com/ordinal-theory-and-the-rise-of-inscriptions}{glassnode.com}.

\bibitem{Helms2023-cg}
K.~Helms, ``Blackrock files for bitcoin trust — analyst calls it a `real
  deal' spot bitcoin {ETF} filing,'' Bitcoin News, 2023, available online from
  \href{https://news.bitcoin.com/blackrock-files-for-bitcoin-trust-analyst-calls-it-a-real-deal-spot-bitcoin-etf-filing}{bitcoin.com}.

\bibitem{ic3-fbi}
{Internet Crime Complaint Center}, ``Alert on cryptocurrency money services
  businesses,'' 2024, \url{https://www.ic3.gov/PSA/2024/PSA240425}.

\bibitem{btcavgtx}
BitInfoCharts, ``Bitcoin avg. transaction fee chart,'' 2025, available online
  from
  \href{https://bitinfocharts.com/comparison/bitcoin-transactionfees.html}{bitinfocharts.com}.

\bibitem{PulpCattel_Wasabi_Observatory}
\BIBentryALTinterwordspacing
{PulpCattel}, ``Wasabi observatory: A list of statistics of the wasabi
  wallet,'' GitHub, 2020. [Online]. Available:
  \url{https://github.com/PulpCattel/Wasabi_Observatory}
\BIBentrySTDinterwordspacing

\bibitem{glaeser2022foundations}
N.~Glaeser, M.~Maffei, G.~Malavolta, P.~Moreno-Sanchez, E.~Tairi, and S.~A.~K.
  Thyagarajan, ``Foundations of coin mixing services,'' in \emph{ACM CCS'22},
  2022, pp. 1259--1273.

\bibitem{Heilman2017-nk}
E.~Heilman, L.~AlShenibr, F.~Baldimtsi, A.~Scafuro, and S.~Goldberg,
  ``{TumbleBit}: An untrusted bitcoin-compatible anonymous payment hub,'' in
  \emph{NDSS'17}.\hskip 1em plus 0.5em minus 0.4em\relax Internet Society,
  2017.

\end{thebibliography}

\end{document}